\documentclass[12pt]{apa}
\usepackage{a4,amssymb,graphicx,eurosym}
\usepackage{apacite}
\usepackage{color}
\makeatletter
\long\def\@makecaption#1#2{%
  \vskip\abovecaptionskip
  \sbox\@tempboxa{#1 #2}%
  \ifdim \wd\@tempboxa >\hsize
    #1 #2\par
  \else
    \global \@minipagefalse
    \hb@xt@\hsize{\hfil\box\@tempboxa\hfil}%
  \fi
  \vskip\belowcaptionskip}
\makeatother
\global\long\def\lb{\bar{\lambda}}
\newcommand{\re}[1]{\color{black} #1 \color{black} }

\begin{document} 

%--------------------------------------------------------------
%	TITLE PAGE
%--------------------------------------------------------------
\begin{center}
{\huge\bf
Spatial statistics and attentional \\[0.7ex]
dynamics in scene viewing}\\
\vspace{10mm}
\large
Ralf Engbert$^{1,2,*}$, Hans A.~Trukenbrod$^{1,2}$, Simon Barthelm\'e$^{2,3}$, and \\
Felix A.~Wichmann$^{2,4-6}$ \\
\vspace{10mm}
\normalsize
$^1$University of Potsdam, Germany\\
$^2$Bernstein Center for Computational Neuroscience Berlin, Germany\\
$^3$University of Geneva, Switzerland\\
$^4$Eberhard Karls University of T\"ubingen, Germany\\
$^5$Bernstein Center for Computational Neuroscience T\"ubingen, Germany\\
$^6$Max Planck Institute for Intelligent Systems, T\"ubingen, Germany\\
\vspace{10mm}
\today
\end{center}

\vspace{10mm}
\noindent
\vspace{\fill}

\noindent
$^*$To whom correspondence should be addressed: \\
Ralf Engbert \\
Cognitive Sciene Program \& Department of Psychology  \\ 
University of Potsdam\\ 
Am Neuen Palais 10\\ 
14469 Potsdam\\
Germany \\ 
E-mail: ralf.engbert\symbol{64}uni-potsdam.de \\
Phone: +49 331 9772140, Fax: +49 331 9772794

\newpage

\newpage
%\pagenumbering{arabic}

%--------------------------------------------------------------
%	ABSTRACT
%--------------------------------------------------------------
\section*{Abstract}
In humans and in foveated animals visual acuity is highly concentrated at the center of gaze, so that choosing where to look next is an important example of online, rapid decision making. Computational neuroscientists have developed biologically-inspired models of visual attention, termed saliency maps, which successfully predict where people fixate on average. Using point process theory for spatial statistics, we show that scanpaths contain, however, important statistical structure, such as spatial clustering on top of distributions of gaze positions. Here we develop a dynamical model of saccadic selection that accurately predicts the distribution of gaze positions as well as spatial clustering along individual scanpaths. Our model relies on, first, activation dynamics via spatially-limited (foveated) access to saliency information, and, second, a leaky memory process controlling the re-inspection of target regions. This theoretical framework models a form of context-dependent decision-making, linking neural dynamics of attention to behavioral gaze data.

%--------------------------------------------------------------
%	INTRODUCTION
%--------------------------------------------------------------
\section*{Introduction}
Research on visual attention models over the past 25 years has resulted in a number of computational models \cite{Borji2013}---using diverse computational mechanisms ---often capable of predicting fixation \emph{locations} for a given input image with reasonable accuracy  \cite{Itti1998,Kienzle2009,Torralba2006,Tsotsos1995}. The models compute so-called \emph{saliency maps}, highlighting those parts of an input image that stand out relative to the surrounding areas \cite{Itti2001}. However, the human visual system is foveated, i.e., it is only able to acquire high-resolution information from a very limited region surrounding the current gaze position (the fovea). Outside the foveal region, visual acuity falls off rapidly, while the effects of visual crowding increase, so that visual processing in the periphery has very limited resolution \re{\cite{Jones1947,Levi2008,Rosenholtz2012}.} 

As a consequence, to explore an entire visual scene we must shift our gaze continually to new regions of interest by producing rapid eye movements (saccades) about three to four times per second \cite{Findlay2012}. Thus, given the progress on mathematical models of visual attention, there is an increasing need for computational models that bridge the gap between static saliency maps---which a human observer's visual system can only know \emph{after} exploring the entire image with its fovea---and the dynamical principles of saccadic selection underlying the generation of scanpaths by human observers. Moreover, part of the mismatch between computer-generated saliency maps and actual gaze patterns might be explained by properties of the visuomotor system \cite{Findlay1999}. \re{Recently, a number of publications addressed specific aspects of this problem, e.g., different roles for short and long saccades \cite{Tatler2006} or return saccades \cite{Ludwig2009,Wilming2013}. Moreover, behavioral biases might produce an important contribution to eye-movement statistics \cite{Tatler2009}. What is currently missing is an integrative computational model that addresses the key aspects of visuomotor control in a coherent theoretical framework. We set out to develop one possible integrative model. }

Spatial patterns of gaze positions carry rich information on the processes of saccadic selection by the human visual system, and this information can be analyzed applying methods from the theory of spatial point processes \cite{Illian2008,Barthelme2013}. Saliency maps aim at the prediction of two-dimensional (2D) densities of gaze patterns (first-order spatial statistics). However, saliency maps do not contain the rich information about spatial interactions inherent in experimental eye-tracking data: fixations are interdependent. Second-order statistics provide quantitative tools to investigate interactions in gaze patterns. \re{Such interactions} in turn may be used to gain information about the processes \cite{Law2009} underlying the generation of neigboring gaze positions, which themselves are directly related to models of saccadic selection. 

We start with analyzing the spatial statistics of gaze patterns using point process theory \cite{Illian2008} and show that gaze patterns are characterized by small-scale clustering, in addition to the inhibition-of-return mechanism \cite{Klein2000} that is thought to represent the dominant dynamical principle in extant attention models \cite{Itti2001}. Next, since  these results provide strong constraints for possible neural mechanisms of saccadic selection, we develop a dynamical model for real-time attention allocation and gaze control based on activation-based maps \cite{Engbert2011,Engbert2012}. \re{Finally,} the model is compared against a range of statistical null models using methods of spatial statistics.

%-----------------------------
%  Methods
%-----------------------------
\section*{Methods}
\subsection*{Experiment}
{\sl Stimulus material}: A set of \re{30} randomly selected, natural landscape photographs (color) was presented to human observers on a 20\verb+"+ CRT monitor (Mitsubishi Diamond Pro 2070; frame rate \re{120}~Hz; resolution: 1280$\times$1024 pixels). Images were classified into two categories, natural \re{object-based scenes (image set 1: 15 images)} vs.~\re{images showing} abstract \re{natural patterns (image set 2: 15 images).} \re{All images } were presented centrally with gray borders extending 32 pixels to the top/bottom and 40 pixels to the left/right of the image, since accuracy of eye tracking systems falls off towards the monitor edges.\\[1ex]
{\sl Task and procedure}: Participants were instructed to position their heads on a chin-rest in front of a computer screen at a viewing distance of 70~cm. Eye movements were recorded binocularly using an Eyelink 1000 video-based eye-tracker (SR-Research, Osgoode/ON, Canada) with a sampling rate of 1000~Hz. Trials began with a black fixation cross presented on gray background at a random location within the image boundaries. After successful fixation, the fixation cross was replaced by the image for 10~s. Participants were instructed to explore each scene for a subsequent memory test. During the experiment, we presented 30 images twice.  Here we limit our analysis to the first presentation of natural landscape photographs.\\[1ex]
{\sl Participants}: We recorded eye movements from 35 participants (20 female, 15 male) aged between 17 and 36 years (mean age: 24 years) with normal or corrected-to-normal vision. Participants were recruited from the University of Potsdam and from a local school (32 students, 3 pupils). All participants received credit points or 8\euro~for participation.\\[1ex]
{\sl Data preprocessing and saccade detection:} We applied a velocity-based algorithm for saccade detection \cite{Engbert2003,Engbert2006}. Saccades had a minimum amplitude of 0.5$^\circ$ and exceeded the average velocity during a trial by 6 standard deviations for at least 6~ms. Eye traces between two successive saccades were tagged as fixations with a mean fixation position \re{averaged} across both eyes. Since eye position was determined by the presentation of a fixation cross at the beginning of a trial, we excluded all \re{initial} fixations from the data set \re{(image set 1: 525; image set 2: 525).} Furthermore, we removed fixations containing a blink or with a blink during an adjacent saccade \re{(image set 1: 580; image set 2: 588).} Overall, \re{the number of} fixations remaining for further analyses was 13,349 \re{(image set 1) and 12,740 (image set 2).}

\subsection*{Spatial statistics}
Gaze positions can be interpreted as realizations from a spatial point process \cite{Illian2008} that can be represented as the random set of points $N=\{x_1,\,x_2,\,x_3,\,...\}$ (also called a point pattern). The 2D density (or intensity) $\lambda$ of the spatial point process is given as the expectation or mean value of the number of points in an observation window $B$, i.e., $\lambda=E(n(B))$, where $n(.)$ is a counting measure. A process is statistically homogeneous (or stationary), if $N$ and the translated set $N_x=\{x_1+x,\,x_2+x,\,x_3+x,\,...\}$ have the same distribution for all $x$. For a stationary spatial point process, the intensity $\lambda$ is constant over space. For a non-stationary process, the intensity is a function of location, $\lambda=\lambda(x)$. For the computation of densities from experimental data, we used kernel-density estimates with bandwidth parameters chosen according to Scott's rule \cite{SpatStat,Scott1992}. To compute deviations between 2D densities $P_{kl}$ and $Q_{kl}$ \re{at grid position $(k,l)$}, we used the Kullback Leibler divergence, a symmetric version of the information gain \cite{Beck1993}, i.e.,
\begin{equation}
\Delta_{KLD} = \frac{1}{2}\sum_{k,l}\left(P_{kl}\log\frac{P_{kl}}{Q_{kl}} + Q_{kl}\log\frac{Q_{kl}}{P_{kl}} \right) \;.
\end{equation}

Second-order statistics \re{(see also the illustrated notes in the Appendix)} are based on the pair density $\rho(x_1,x_2)$, which gives the probability $\rho(x_1,x_2){\rm d}x_1{\rm d}x_2$ of observing points in each of two disks $b_1$ and $b_2$ with linear dimensions ${\rm d}x_1$ and ${\rm d}x_2$, respectively. Point patterns can be characterized by the pair density, which is typically a function of the pair distance, i.e., $\rho(x_1,x_2)=\rho(r)$ with $r=\|x_1-x_2\|$, for two arbitrary realizations $x_1$ and $x_2$. Using a kernel-based method, a estimator for the pair density can be written as
\begin{equation}
\hat{\rho}(r) = \sum_{x_1,x_2\in W}^{\neq} \frac{k(\|x_1-x_2\|-r)}{2\pi rA_{\|x_1-x_2\|}} \;,
\end{equation}
where $k(.)$ is \re{an} appropriate kernel and \re{$A_{\|x_1-x_2\|}$} denotes an edge correction at distance \re{$\|x_1-x_2\|$} \re{\cite{SpatStat}. For numerical computations we used the Epanechnikov kernel \cite{Illian2008}, i.e.,
\begin{equation}
k(x) = \left\{ \begin{array}{cl} \frac{3}{4h}(1-\frac{x^2}{h^2}), \; &  \mbox{for}\; -h\le x\le h \\
 0, & \mbox{otherwise}
 \end{array} \right.
\end{equation}
The problem of choosing the bandwidth $h$ appropriately is frequently discussed in the literature \cite{Illian2008}. The bottom line from this discussion is that the behavior of the estimator should be analyzed over a range of bandwidths. We will run such an analysis below (Fig.~2).}

The pair correlation function $g(r)$ is \re{a} normalization of the pair density with respect to first-order intensity $\hat{\lambda}$, so that the estimator for the pair correlation is given by $\hat{g}(r) = \rho(r)/\hat{\lambda}^2$. The interpretation of the pair correlation function for a given point pattern is straightforward. For a random pattern without clustering, the pair correlation function is $\hat{g}(r) \approx 1$ across the full range of distances $r$. If $\hat{g}(r)>1$, then pairs of fixations are more abundant than on average at distance $r$. If $\hat{g}(r)<1$, then pairs of fixations are less abundant than on average at a distance $r$. Thus, the pair correlation function $\hat{g}(r)$ measures how selection of a particular point location (i.e., fixation position) is influenced by other fixations at distance $r$.

Using the inhomogeneous pair correlation function $g_{inhom}(r)$, we can remove the first-order inhomogeneity from the second-order spatial statistics, i.e.,
\begin{equation}\label{Eq_ginhomdef}
\hat{g}_{inhom}(r) = \sum_{x_1,x_2\in W}^{\neq} 
\frac{1}{\hat{\lambda}(x_1)\hat{\lambda}(x_2)}
\frac{k(\|x_1-x_2\|-r)}{2\pi rA_{\|x_1-x_2\|}} \;.
\end{equation}
Estimation of $\hat{g}_{inhom}(r)$ involves two steps: First, we estimated the overall intensity $\hat{\lambda}(x)$ for all fixation positions obtained for a given scene. In this procedure we borrow strength from the full set of observations to obtain reliable estimates of the inhomogeneity. Second, we computed the pair correlation function \re{from a single trial with respect to the inhomogeneous density of the full data set.}

In case of a given pair correlation function $\hat{g}(r)$, the scalar quantity \cite{Illian2008}
\begin{equation}\label{Eq_PCFdev}
\Delta_g = \int_0^\infty (\hat{g}(r)-1)^2 dr \;,
\end{equation}
denoted as PCF deviation in the following, serves as a useful test statistic that quantifies the deviations from randomness for a given point pattern with inhomogeneous density $\hat{\lambda}(x)$. The integral in Eq.~(\ref{Eq_PCFdev}) was evaluated numerically for pair distances $r$ between 0.1$^\circ$ and 5$^\circ$ \re{(image set 1 and 2) and between 0.1$^\circ$ and 3$^\circ$ (Le Meur et al.~data, see below).}

%---------------------------------------------------------------
%	RESULTS
%----------------------------------------------------------------
\section*{Results}
We conducted an eye-tracking experiment on scene viewing with \re{35} human observers using \re{15 object-based} natural  scenes \re{(image set 1).} Resulting gaze data were evaluated using first- and second-order spatial statistics; we found that data exhibit unexpected spatial \re{aggregation (or} clustering). \re{We reproduced this finding for a set of 15 abstract natural patterns (image set 2) and for an external dataset \cite{LeMeur2006} that was made publicly available \cite{mit-saliency-benchmark}.} Based on \re{these results,} we developed a dynamical model for saccadic selection that was evaluated by the spatial-statistics approach introduced in this section.

%---------------------------------------------------------------
%	Spatial statistics
%---------------------------------------------------------------
\subsection*{Spatial statistics and pair correlation function}
We began by numerically computing the spatial (2D) density of gaze positions from experimental data \re{collected for image set 1} ({Fig.~1a}). Fixation positions are indicated by red dots (a total of 930 fixations from 35 observers for image \#2). Densities were computed using a 2D kernel density estimator \cite{Rlang,SpatStat} (see Methods) and are visualized by gray shading in the plot. The bandwidth parameter $h_{\rm density}$ for the kernel density estimation was computed according to Scott's rule \cite{Scott1992} (range from 1.8$^\circ$ to 2.2$^\circ$ for $h$ over the full set of 15 images). The obtained 2D density $\hat{\lambda}(x,y)$ is inhomogeneous because of the dependence on position $(x,y)$. A representative sample trajectory from a single trial is given in {Fig.~1d}, where the second and last fixation of the scanpath \re{is} highlighted by white color and by their serial numbers. The first fixation was omitted, since all trials started at \re{a random position within an image determined by} our experimental procedure (see Methods).

%--------
% Fig. 1
%--------

The pair correlation function $g(r)$ gives a quantitative summary of interactions in fixation patterns by measuring how distance patterns between fixations differ from what we would expect from independently distributed data (see Appendix). A value of $g(r)$ above 1 for a particular \re{distance} $r$ indicates clustering, meaning that there are more pairs of points separated by a distance $r$ than we would expect if fixation locations were statistically independent.  

For the estimation of the pair correlation function, short sequences of gaze positions from single experimental trials were considered. However, spatial inhomogeneity of the 2D density was taken into account. To obtain a reliable estimate of the spatial inhomogeneity, the 2D density was estimated from the full data-set ({Fig.~1a}) of all fixations on a given image taken from all participants and trials. It is important to note, however, that in the computation of the kernel density estimate $\hat{\lambda}(x)$ used for the inhomogeneous pair correlation function, Eq.~(\ref{Eq_ginhomdef}), an optimal bandwidth parameter $h$ is needed to avoid two possible artifacts: First, if $h$ is very small, then spatial correlations might be underestimated due to overfitting of the inhomogeneity of the density. Second, if $h$ is too large, then spatial correlations might be overestimated, since first-order inhomogeneity is not adequately removed from the second-order spatial statistics. We solved this problem by computing the PCF deviation $\Delta_g$ for the inhomogeneous point process for varying values of the bandwidth $h$ ({Fig.~\ref{Fig_BW-PCF}}). \re{Since the inhomogeneous point process generates uncorrelated fixations (i.e., $g_{\rm theo}(r)=1$), the optimal bandwidth for the dataset corresponds to a minimum of the PCF deviation $\Delta_g$ (quantifying the deviation from the ideal value $g(r)=1$). For image set 1, the optimal value was estimated as  $\hat{h}_1=4.0^\circ$ ({Fig.~\ref{Fig_BW-PCF}}a).}

%--------------
% Figure 2
%--------------

Based on this density estimate, we can compute the inhomogeneous pair correlation function $g_{inhom}(r)$, in which first-order inhomogeneity is removed from the second-order spatial correlations (see Methods). As a result, we obtained pair correlations from individual trials ({Fig.~1g}, gray lines). Deviations from $g_{inhom}(r)\approx 1$ indicate spatial clustering at a specific distance $r$. The mean pair correlation function $\bar{g}_{inhom}(r)$ provides evidence for clustering at small spatial scales with $r<4^\circ$ ({Fig.~1g}, red line). Such a scale is greater than the foveal zone $(r<2^\circ)$ and might provide an estimate of the size of the {\sl effective} perceptual window in free scene viewing. \re{This result is compatible with earlier findings that the zone of active selection of saccade targets extends beyond the fovea into the parafovea up to eccentricities of $4^\circ$ \cite{Reinagel1999}.}

Next, we carried out the same numerical computations for two sets of surrogate data. The surrogate data were generated to test the null hypotheses of complete spatial randomness, both for an inhomogeneous point process with position-dependent intensity $\lambda(x,y)$ and for a homogeneous point process with constant intensity $\lambda_0$. For the inhomogeneous point process, we sampled from the estimated intensity $\hat{\lambda}(x,y)$ ({Fig.~1b}), whereas a constant intensity $\hat{\lambda}_0$, obtained from spatial averaging, was used for the homogeneous point process ({Fig.~1c}). Both surrogate datasets are important for checking the reliability of the computation of the pair correlation function for the original data ({Fig.~1g}). First, the inhomogeneous point process gives a flat mean correlation function with $g(r)\approx 1$ ({Fig.~1h}), which demonstrates the absence of clustering (except for the divergence at very small scales as an effect of numerical computation issues). Thus, the spatial correlations in the experimental data \re{are} not a simple consequence of spatial inhomogeneity. Second, the result for the homogeneous point process ({Fig.~1h}) is the same as for the inhomogeneous point process, which indicates that the correction for inhomogeneity needed for computations in {Fig.~1g} does not produce unwanted artefacts due to possible overfitting of the spatial inhomogeneities. We conclude that our experimental data give a clear indication for spatial clustering at length-scales smaller than $4^\circ$ of visual angle. Additionally, we checked the hypothesis that this effect of spatial clustering might be due to saccadic undershoot and subsequent short correction saccades by excluding all fixations with durations shorter than 200~ms. A related analysis of the PCF indicates no qualitative differences from the original data ({Fig.~1h}).

Results reported so far were obtained for a single, representative image. Over the full set of 15 images, we analyzed \re{the 2D densities using} a symmetrized form of the Kullback-Leibler divergence (KLD) based on the concept of information gain \cite{Beck1993}. For the experimental data, we applied a split-half procedure (first half of participants vs.~second half of participants) and computed the KLD between the two experimental densities. The corresponding KLD values demonstrate that the inhomogeneous point process reproduces the 2D density ({Fig.~1f}), while the homogeneous point process clearly fails to approximate the systematic inhomogeneity in the image. Model type had a significant effect on KLD, $\chi^2(2)=105.4$, $p<0.01$. Contrasts revealed that (1) spatially inhomogeneous data-sets were different from the homogeneous data, $b=0.319$, $t(28)=22.85$, $p<0.01$, and that (2) experimental data were significantly different from inhomogeneous surrogate data, $b=0.062$, $t(28)=2.56$, $p=.016$.

Both sets of surrogate data were submitted to an analysis of the pair correlation function, where the deviations from $g(r)\approx 1$ were computed to obtain a PCF measure indicating the amount of spatial correlation averaged over distances (see Methods). Results indicate that the surrogate data produce---as designed---uncorrelated gaze positions (low PCF deviation), while the experimental data by human observers exhibit spatially correlated gaze positions ({Fig.~1i}). Model type had a significant effect on PCF, $\chi^2(2)=98.4$, $p<0.01$. 

%--------------
% Figure 3
%--------------

\re{To investigate the reliability of our finding of spatial clustering on short length scales during natural scene viewing, we investigated two other datasets using the same procedures. We compared the mean pair correlation functions per image for natural object-based scenes (image set 1; Fig.~\ref{Fig_datasets}a,d) and abstract natural patterns (image set 2; Fig.~\ref{Fig_datasets}b,e) with the optimal bandwidths adjusted to the datasets (Fig.~\ref{Fig_BW-PCF}b,c). While variability of the pair correlation function might be greater for the abstract scenes than for the object-based scenes (Fig.~\ref{Fig_datasets}d,e, gray lines) the mean pair correlation functions for the images sets (Fig.~\ref{Fig_datasets}d,e, red lines) are very similar. This result indicates that the clustering on short length scales is a robust phenomenon which does not seem to depend sensitively on scene content. Second, we analyzed a publicly available dataset \cite{LeMeur2006} from the MIT saliency benchmark \cite{mit-saliency-benchmark,Judd2012} consisting of 40 participants who viewed 27 color images for 15 seconds. For these data, the spatial scale of the presentation display for the images (Fig.~\ref{Fig_datasets}c) was considerable smaller than in our experiment. Consequently, we observed pair correlation functions that indicate clustering on smaller scales ($<1.5^\circ$, Fig.~\ref{Fig_datasets}f). Thus, while scene content does not seem to exert a strong influence on spatial correlations, spatial scale of the image modulates the spatial scale of the clustering of fixations. }

We conclude that second-order spatial statistics obtained for the experimental data are significantly different from stochastic processes implementing the assumption of spatial randomness. Furthermore, the mere presence of spatial inhomogeneity in the experimental data cannot explain by itself the observed spatial correlations, which is evident in the results for the inhomogeneous point process. While inhibition-of-return \cite{Klein2000} has been discussed frequently as one of the key principles added to saliency maps for saccadic selection \cite{Itti2001}, spatial clustering of gaze positions is an additional statistical property that is highly informative on mechanisms of gaze planning, but has been neglected so far. Next, we use these results to develop and test a dynamical model for saccade generation that uses activation field dynamics to reproduce spatial statistics of first- and second-order.

%--------------------------------------------------------------
%	Model
%--------------------------------------------------------------
\section*{A dynamical model of saccade generation}
A key assumption for the model we propose is the combination of two neural activation maps to implement dynamical principles for saccadic selection. First, a fixation map $f(x,y;t)$ is keeping track of the sequence of fixations by inhibitory tagging \cite{Itti2001}. Second, an attention map $a(x,y;t)$ that is driven by early visual processing controls the distribution of attention. \re{Physiologically,} the assumption of the dynamical maps is supported by the presence of an allocentric motor map of visual space in the primate entorhinal cortex \cite{Killian2012}. Moreover, this map is spatially discrete \cite{Stensola2012} and serves as a biological motivation for the fixation and attention maps in our model.

We implemented activation maps for attention and fixation (inhibitory tagging) on a discrete square lattice of dimension $L\times L$. Lattice points $(i,j)$ have equidistant spatial positions $(x_i,y_j)$ for $i,\,j = 1,\,...,\,L$, where $x_i=x_0+i\Delta x$ and $y_j=y_0+j\Delta y$. As a consequence, attention and fixation maps are implemented in spatially discrete forms, $\{a_{ij}(t)\}$ and $\{f_{ij}(t)\}$, respectively. For the numerical simulations, time was discretized in steps of $\Delta t=10$~ms with $t=k\cdot\Delta t$ and $k=0,\,1,\,2,\,...,\,T$. 
  
If the observer's gaze is at position $(x_g,y_g)$ at time $t$, then a position-dependent activation change $F_{ij}(x_g,y_g)$ and a global decay proportional to the current activation $-\omega f_{ij}(t)$ are added to all lattice positions to update the activation map at time $t+1$, i.e.,
\begin{equation}
\label{Eq_fixmap}
f_{ij}(t+1) = F_{ij}(x_g,y_g) + (1-\omega)f_{ij}(t) \;,
\end{equation}
where the activation change $F_{ij}(x_g,y_g)\equiv F_{ij}(t)$ is implicitly time-dependent because of the time-dependence of gaze positions $(x_g(t),y_g(t))$.  The constant $\omega\ll 1$ determines the strength of the decay of activation. For the spatial distribution of the activation change $F_{ij}(t)$ we assume a Gaussian profile, i.e.,  
\begin{equation}
\label{Eq_gaussian}
F_{ij}(t) = \frac{R_0}{\sqrt{2\pi}\sigma_0}
\exp\left(-\frac{(x_i-x_g(t))^2 + (y_j-y_g(t))^2}{2\sigma_0^2}\right) \;,
\end{equation} 
with the free parameters $\sigma_0$ and $R_0$ controlling the spatial extent of the activation change and the strength of the activation change, respectively. In our model, the build-up of activation in the fixation map is a mechanism of inhibitory tagging \cite{Itti2001} to reduce the amount of refixations on recently visited image patches.

For the attention map $a_{ij}(t)$ we assume similar dynamics, however, the width of Gaussian activation change $A_{ij}(t)$ is assumed to be proportional to the static saliency map $\{\phi_{ij}\}$. The updating rule for the attention map is given by
\begin{equation}
\label{Eq_attmap}
a_{ij}(t+1) = \frac{\phi_{ij}A_{ij}(t)}{\sum_{kl}\phi_{kl}A_{kl}(t)} + (1-\rho)a_{ij}(t) \;,
\end{equation}
with decay constant $\rho\ll 1$. As a result, the saliency map $\phi_{ij}$ is accessed locally through a Gaussian aperture with size $\sigma_1$ and scale parameter $R_1$, similar to Eq.~(\ref{Eq_gaussian}). Using the local read-out mechanism, information is provided for the attention map to identify regions of interest for eye guidance. 

The fixation map monitors recently visited fixation locations by increasing local activation at the corresponding lattice points \cite{Engbert2011,Freund1992}. If the observer's gaze position is at a position corresponding to lattice position $(i,j)$, then a position-dependent activation change $F_{ij}$ in the form of a Gaussian profile is added locally in each time step, while a global decay proportional to the current activation is applied to all lattice positions. The width of the Gaussian activation $\sigma_0$ and the decay $\omega$ are the two free parameters controlling activation in the fixation map. For the attention map $a_{ij}(t)$ we assume similar dynamics, including local increase of activation with size $\sigma_1$ and global decay $\rho$. However, the amount of activation change $A_{ij}$ is assumed to be proportional to the time-independent saliency map $\phi_{ij}$, so \re{that} the local increase of activation is $+\phi_{ij}A_{ij}/\sum_{kl}A_{kl}$. 

Our modeling assumptions are related to specific hypotheses on model parameters. We expect that the size of the Gaussian profile for the attention map is larger than the corresponding size of the fixation map, $\sigma_1>\sigma_0$, since attention is the process driving eye movements into new regions of visual space, while the inhibitory tagging process should be more localized. A similar expectation can be formulated on the decay constants. Since inhibitory tagging is needed on a longer time scale as a foraging facilitator, we expect a slower decay in the fixation map compared to the attention map, i.e., $\omega<\rho$.

Next, we assume that, given a saccade command at time $t$, both maps are evaluated to select the next saccade target. First, we apply a normalization of both attention and fixation maps as a general neural principle to obtain relative activations \cite{Carandini2011}. Second, we introduce a potential function as the difference of the normalized maps, 
\begin{equation}\label{Eq_Pot}
u_{ij}(t) = -\frac{[a_{ij}(t)]^{\lambda}}{\sum_{kl}[a_{kl}(t)]^{\lambda}} 
+ \frac{[f_{ij}(t)]^{\gamma}}{\sum_{kl}[f_{kl}(t)]^{\gamma}}  \;,
\end{equation}
where the exponents $\lambda$ and $\gamma$ are free parameters. However, a value of $\lambda=1$ is a necessary boundary condition to obtain a model that accurately reproduces the densities of gaze positions. In a qualitative analysis of the model (see Appendix), pilot simulations showed that $\gamma$ is an important control parameter determining spatial correlations, where $\gamma\approx 0.3$ was used to reproduce spatial correlations observed in \re{our} experimental data. 

The potential $u_{ij}(t)$, Eq.~(\ref{Eq_Pot}), can be positive or negative at position $(i,j)$. Lattice positions with a positive potential, $u_{ij}>0$, are excluded from saccadic selection, since corresponding regions \re{were} visited recently with high probability. Among the lattice positions with negative activations, we implemented stochastic selection of saccade targets proportional to relative activations, also known as Luce's choice rule \cite{Luce1959}. We implemented \re{this form of} stochastic selection from the set ${\cal S}=\{(i,j)|u_{ij}<0\}$, where the probability $\pi_{ij}(t)$ to select lattice position $(i,j)$ at time $t$ as the next saccade target is given by
\begin{equation}\label{Eq_saccsel}
\pi_{ij}(t) = \max\left(\frac{u_{ij}(t)}{\sum_{(k,l)\in {\cal S}} u_{kl}(t)},\,\eta\right) \;.
\end{equation}
The noise term $\eta$ is an additional parameter controlling the amount of noise in target selection.

%----------------------------------------------------------------
%	Numerical simulations and results
%----------------------------------------------------------------
\subsection*{Numerical simulations of the model}
Our computational modeling approach to saccadic selection has been developed to propose a minimal model that captures the types of spatial statistics observed in experimental data. \re{After fixing model parameters, all that is needed to run the model on a particular image is a 2D density estimate of gaze patterns (from experimental data) or a corresponding 2D prediction from one of the available saliency models \cite{Borji2013}.} To reduce computational complexity \re{in the current study and to exclude potential mismatches between data and model simulations due to the saliency models,} we run the model on experimentally realized densities of gaze positions, which is equivalent to assuming an exact saliency model. However, our modeling approach is compatible with future dynamical saliency models that provide time- and position-dependent saliency during a sequence of gaze shifts, thus our model introduces a general dynamical framework and is not tied to using \re{empirical data}. 

The numerical values of the 5 model parameters were estimated from experimental data recorded for the first 5 images \re{of natural object-based scenes (images set 1) }using a genetic algorithm approach \cite{Mitchell1998} (see Appendix, Table 1). \re{The remaining 10 images of image set 1 were used for model evaluations---as the 15 images of image set 2.} The objective function for parameter estimation was based on evaluation of first-order statistics (2D density of gaze positions) and the distribution of saccade lengths. In agreement with our first expectation, the estimated optimal values for the spatial extent of the inhibitory tagging process in the fixation map, $\sigma_0=2.2^\circ$, is considerably smaller than the corresponding size of the build-up function for the attention map, $\sigma_1=4.9^\circ$. Our second expectation was related to the decay constants, which turned out to be larger for the attention map, $\rho=0.066$, than for the fixation map, $\omega=9.3\cdot10^{-5}$, so $\rho$ was greater than $\omega$, again as expected. Finally, the noise level in the target map is $\eta=9.1\cdot10^{-5}$.

%-------------
% Figure 4
%-------------

%--------------
% Figure 5
%--------------

An example for the simulation of the model demonstrates the interplay between inhibitory processes from the fixation map and the attention map during gaze planning ({Fig.~\ref{Fig_simexam}}). The fixation map builds up activation at fixated lattice positions (yellow to red), while the attention map identifies new regions of interest for saccadic selection (blue). These simulations show on a qualitative level how the model implements the interplay of the assumed mechanisms of inhibitory tagging and saccadic selection of gaze positions (see {Supplementary Video}).

To investigate model performance qualitatively, we ran simulations for one image (image \#6, 930 fixation, {Fig.~\ref{Fig_modelanalysis}a}) and obtained a number of fixations similar to the experimental data (882 fixations, {Fig.~\ref{Fig_modelanalysis}b}). Single-trial scanpaths from experiments and simulations are shown additionally in {Fig.~\ref{Fig_modelanalysis}a,b}). The resulting distributions of saccade lengths indicate that our dynamical model is in good agreement with experimental data, while the two surrogate datasets (homogeneous and inhomogeneous point processes) fail to reproduce the distribution ({Fig.~\ref{Fig_modelanalysis}c}). An analysis of the pair correlation functions indicates that the spatial correlations present in the experimental data were approximated by the dynamical model ({Fig.~\ref{Fig_modelanalysis}d}), however, the two surrogate datasets representing uncorrelated sequences by construction produce qualitatively different spatial correlations.

To investigate the influences of saccade-length distributions on pair correlations, we constructed \re{another} statistical control model that had access to the image-specific saccade-length \re{distribution. It is important to note that} this model was not introduced as a competitor to the dynamical model, which is able to {\sl predict} saccade-length distributions. The statistical control model approximated the distribution of saccade lengths $l$ and 2D densities of gaze positions $x$ by sampling from the joint probability distribution $p(x,l)$ under the assumption of statistical independence of saccade lengths and gaze positions, i.e., $p(x,l)=p(x)p(l)$. This model, by construction, approximates the distribution of saccade lengths and 2D density of gaze positions ({Fig.~\ref{Fig_modelanalysis}c}). The simulations indicate, however, that even the combination of inhomogeneous density of gaze positions and non-normal distribution of saccade lengths used by the statistical control model cannot explain spatial correlations in the experimental data characterized by the pair correlations function ({Fig.~\ref{Fig_modelanalysis}d}).

For the statistical analysis of model performance on new images, we carried out additional numerical simulations. We fitted model parameters to data obtained for the first 5 images only (see above) and predicted data for the remaining 10 images by the new simulations to isolate parameter estimation from model evaluation (calculating test errors rather than training errors).

%-------------
% Figure 6
%-------------

Our simulations show that the dynamical model predicted the 2D density of gaze positions accurately (Fig.~\ref{Fig_ModelResults}a). The  obtained KLD values for the model (blue) were comparable to KLD values calculated by the split-half procedure for the experimental data (red) and to the KLD values obtained for the statistical control models (green=Homogeneous point process, yellow=Inhomogeneous point process, magenta=Control model). Model type had a significant effect on KLD, $\chi^2(4)=109.4$, $p<0.01$. Posthoc comparisons indicated significant effects between all models ($p<0.01$) except for the comparison between experimental data and the dynamical model ($p=.298$) and for the comparison between dynamical model and inhomogeneous point process ($p=.679$). In an analysis of the PCF estimated from the same set of simulated data (Fig.~\ref{Fig_ModelResults}b), the dynamical model (blue) produced deviations from an uncorrelated point process that are in good agreement with the experimental data (red). Model type had a significant effect on PCF deviation, $\chi^2(4)=91.6$, $p<0.01$. Posthoc comparisons indicated significant effects for all comparisons ($p<0.01$) except for the comparison between experimental data and the dynamical model ($p=.990$) and between homogeneous and inhomogeneous point processes ($p=.996$).

\re{To check the reliability of the results, we performed all corresponding calculations for the images showing abstract natural scenes (image set 2). For these simulations, we used the set of model parameters fitted to the first 5 images of image set 1, which corresponds to the hypothesis that scene content (object-based scenes vs.~abstract natural patterns) does not have a strong impact on spatial correlations in the scanpath data. For the simulated densities (Fig.~\ref{Fig_ModelResults}c), we reproduced the statistical results from image set 1. Again, model type had a significant effect on KLD, $\chi^2(4)=162.6$, $p<0.01$. Posthoc comparisons indicated significant effects between all models ($p<0.01$) except for the comparison between experimental data and the dynamical model ($p=.251$), for the comparison between dynamical model and inhomogeneous point process ($p=.784$), and for the comparison between inhomogeneous point process and experimental data ($p=0.012$). For the pair correlation function (Fig.~\ref{Fig_ModelResults}d), model type had a significant effect ($\chi^2(4)=140.6$, $p<0.01$) and posthoc comparisons indicated significant effects for all comparisons ($p<0.01$) except for the comparison between experiment and dynamical model ($p=.453$) and between homogeneous and inhomogeneous point processes ($p=.700$). Thus the main statistical results obtained from image set 1 were reproduced for images of abstract natural patterns (image set 2). These results lend support to the hypothesis that scene content does \re{not} have a strong influence \re{on} second-order spatial statistics of gaze patterns.}

Thus, \re{the} dynamical model performed better than any of the statistical models in predicting the average pair correlations. Although \re{one of our} statistical control models generated data by using image-specific saccade-length information in addition to the 2D density of gaze position, it could not predict the spatial correlations as accurately as the dynamical model that was uninformed about the image-specific saccade-length distribution.

%-------------------------------------------------------
%	DISCUSSION
%-------------------------------------------------------
\section*{Discussion}
Current theoretical models of visual attention allocation in natural scenes are limited to the prediction of first-order spatial statistics (2D densities) of gaze patterns. We were interested in attentional dynamics that can be characterized by spatial interactions (as found in the second-order statistics). Using the theory of spatial point processes, we discovered that gaze patterns can be characterized by clustering at small length scales, which cannot be explained by spatial inhomogeneity of the 2D density. We proposed and analyzed a model based on dynamical activation maps for attentional selection and inhibitory control of gaze positions. The model reproduced 2D densities of gaze maps (first-order statistics) and distributions of saccade lengths as well as pair correlations (second-order spatial statistics). 

\subsection*{Spatial statistics}
While research on the computation of visual saliency has been a highly active field of research \cite{Borji2013}, there is currently a lack of computational models for the generation of scanpaths on the basis of known saliency. Inhibition-of-return \cite{Klein2000} has been proposed as a key principle to prevent continuing refixation within regions of highest saliency. However, our analysis of the pair correlation function demonstrates that saccadic selection at small length scales is dominated by spatial clustering. \re{Thus, our findings are highly compatible with the view that inhibition-of-return cannot easily be observed in eye-movement behavior in natural scene viewing experiments \cite{Smith2011}. However, the} spatial correlations can be exploited to investigate dynamical rules underlying attentional processing in the visual system. Our experimental data show a clear effect of spatial clustering for length scales shorter than about $4^\circ$. \re{However, these results are incompatible with} the current theory of saliency-based attention allocation combined with an inhibition-of-return \re{mechanism. Future simulations will investigate the predictive power of the model when saliency models are used as input.}  

\subsection*{Modeling spatial correlations}
In a biologically plausible computational model of saccade generation, a limited perceptual span needs to be implemented for attentional selection \cite{Findlay2012}. We addressed this problem by assuming a Gaussian read-out mechanism with local retrieval from the saliency map through a limited aperture, analogous to the limited extend of high-fidelity information uptake through the fovea. We used the experimentally observed density of fixations as a proxy for visual saliency. 

First, our results indicate that a very limited attentional span (Gaussian with standard deviation parameter $\sim 4.9^\circ$) of about twice the size of the activation mechanism for tracking the gaze positions ($\sim 2.2^\circ$) is sufficient for saccade planning. This attentional span is efficient, however, since the combination of fixation and attention maps in our model actively drives the model's gaze position to new salient regions computed via normalization of activations \cite{Carandini2011}.

Second, our model correctly predicted spatial clustering of gaze positions at small length scales. The pair correlation function indicates that there is a pronounced contribution by refixations very close to the current gaze position. This effect is compatible with the distribution of saccade lengths, however, a statistical control model that generated data from statistically independent probabilities of 2D density and saccade lengths could not reproduce the pair correlations adequately. 

\re{\subsection*{Implications for computational models of active vision}
During active vision, our visual system relies on frequent gaze shifts to optimize retinal input. Using a second-order statistical analysis we demonstrated that spatial correlations across scanpaths might provide important constraints for computational models of eye guidance. In our dynamical model, spatial clustering at small scales is the result of two principles. First, the fixation map is driven  by an activation function with a small spatial extent ($\sim 2.5^\circ$). Second, the time-scale of activation build-up in the fixation map is slow compared to the build-up of activation in the  attention map. Both mechanisms permit refixations at positions very close to the current gaze position before the system moves on to new regions of visual space.
}

\subsection*{Limitations of the current approach}
The current work focused on spatial statistics of gaze patterns and we \re{proposed and analysed} dynamical mechanisms of eye guidance in scene viewing. In our model, a Gaussian read-out mechanism for the static empirical saliency map was implemented as a simplification. A more biologically plausible combination of our model of eye guidance with a dynamical saliency model \re{\cite{Borji2013}} is a natural extension of the current framework, and the development of such a model is work in progress in our laboratories. Clearly, the current modeling architecture is not limited to input from static saliency maps.

Another simplification is related to the timing of saccades \cite{Nuthmann2010}. In the current version of our model, we implemented random timing and sampled fixation durations randomly from a predefined distribution. More adequate models of fixation durations, however, will need to include interactions of processing difficulty between fovea and periphery \cite{Laubrock2013}. 

%-----------------------------
%  Supplemental Information
%-----------------------------
\subsection*{Supplemental Information}
This work include a supplemental video animation of the model simulations. Experimental data on fixation patterns and computer code for statistical analysis and model simulations will be made available via the Potsdam Mind Research Respository (PMR2, http://read.psych.uni-potsdam.de).  

%-----------------------------
%  Acknowledgments
%-----------------------------
\subsection*{Acknowledgments}
This work was supported by Bundesministerium f\"ur Bildung und Forschung (BMBF) through the Bernstein Computational Neuroscience Programs Berlin (Project B3, FKZ: 01GQ1001F and FKZ: 01GQ1001B to R.E.~and F.A.W., resp.) and T\"ubingen (FKZ: 01GQ1002 to F.A.W.) and by Deutsche Forschungsgemeinschaft (grants EN 471/13--1 and WI 2103/4--1 to R.E. and F.A.W., resp.).

%--------------------------------------------------------
%	REFERENCES AND APPENDICES
%--------------------------------------------------------
\bibliography{Refs}
\bibliographystyle{elsart-harv}

\vspace{10mm}
\begin{appendix} 
\begin{center}
\large Appendix
\end{center}
\subsection{Estimation of model parameters}
Some of the free parameters of the model were set to fixed values to reduce the number of free parameters and to facilitate parameter estimation. First, saccade timing was outside the primary scope of the current work. Time intervals between two decisions for saccadic eye movements are drawn from a gamma distribution of 8th order \cite{Trukenbrod2014} with a mean value of $\mu=275$~ms. Second, we assumed that the build-up of activation is considerably faster in the attention map than in the fixation map by choosing $R_0=0.01$ and $R_1=1$, i.e.,~$R_1/R_0\sim 100$. 

Model parameters were estimated by minimization of a loss function combining information on the densities of gaze positions and of saccade lengths,
\begin{equation}\label{Eq_lossfunc}
\Lambda(\sigma_0,\sigma_1,\omega,\rho,\eta) = 
\sum_i\left(p_i^e - p_i^s\right)^2 + \sum_j\left(q_j^e - q_j^s\right)^2 \;,
\end{equation}
where $p^e$ and $p^s$ are the experimental and simulated distributions of pair distances between all data points for a given image and $q^e$ and $q^s$ are the distributions of saccade lengths for experimental and simulated data, respectively. The minimum of the objective function $\Lambda$ was determined by a genetic algorithm approach (\cite{Mitchell1998}) within a predefined range ({\bf Tab.~\ref{Tab_ModPar}}). Mean values and standard errors of the means were computed from 5 independent runs of the genetic algorithm.

\begin{table}[htdp]
\caption{\label{Tab_ModPar}
Model parameters}
\begin{center}
\begin{tabular}{lcccccc}
\hline\hline
Parameter & Symbol & Mean & Error & Min & Max & Reference \\
\hline
Fixation map\\
\hspace{5mm}Activation span [$^\circ$] & $\sigma_0$ & 
    2.16 & 0.11 & 0.3 & 10.0 & 
    Eq.~(\ref{Eq_gaussian})  \\
\hspace{5mm}Decay	& $\log_{10}\omega$	& 
    $-4.03$ & $0.28$ & $-5.0$ & $-1.0$ 	& 
    Eq.~(\ref{Eq_fixmap})  \\
Attention map \\
\hspace{5mm}Activation span [$^\circ$] & $\sigma_1$ & 
    4.88 	& 0.25 & 0.3 & 10.0 	& 
    Eq.~(\ref{Eq_attmap}) \\
\hspace{5mm}Decay & $\log_{10}\rho$ & 
    $-1.18$ & $0.08$ & $-3.0$ & $-1.0$ & 
    Eq.~(\ref{Eq_attmap})  \\
Target selection \\
\hspace{5mm}Additive noise & $\log_{10}\eta$	& 
    $-4.04$ 	& $0.07$ & $-9.0$ & $-3.0$ & 
    Eq.~(\ref{Eq_saccsel})  \\
\hline
\end{tabular}
\end{center}
\label{table_modelpar}
\end{table}

\subsection{Qualitative analysis of the model}
The pair correlation function was the most important statistical concept in model evaluation. In our model, the strength of spatial correlation turned out to be related to the value of the exponent $\gamma$ in the fixation map of the potential, Eq.~(\ref{Eq_Pot}). We performed numerical simulations with the value of parameter $\gamma$ fixed at different values between 0 and 1 to investigate the dependence of the spatial correlations on this parameter qualitatively ({Fig.~6}). While $\gamma=1$ produces negatively correlated scanpaths, $g(r)<1$, at short pair distances $r$, it is possible to produce even stronger PCF value than in the experimental data for $\gamma<0.3$. Thus, a single parameter in our model can generate a broad range of second-order statistics.

%---------------
% Figure 7
%---------------

\subsection{Some notes on the pair correlation function}
The pair correlation function can
be used to examine the second-order statistics of a point pattern.
We first need to define a few terms. A \emph{point process }is a probability
distribution that generates random point patterns: a sample from a
point process is a set of observed locations (i.e., fixations,
in our case). Therefore, taking two different samples from the same point
process will result in two different sets of locations, although the locations
may be similar ({Fig.~\ref{fig:illustration-intensity}}). 

%------------------
% Figure 8
%------------------

{\sl First-order statistics: the intensity function.}
The first-order statistics of a point process are given by its intensity
function $\lambda\left(x\right)$. The higher the value of $\lambda\left(x\right)$,
the more likely we are to find points around location $x$. {Figure \ref{fig:illustration-intensity}c}
shows the theoretical intensity function for the point process generating
the points in {Figures \ref{fig:illustration-intensity}a} and {\ref{fig:illustration-intensity}b}.

One way to look at the first-order statistics of a point process is
via random variables that count how many points fall in a given region.
For example, we could define a variable $c_{A}$ that counts how many
points fall within area $A$, for a given realisation of the point
process. The expectation of $c_{A}$ (how many points fall in $A$
on average) is given by the intensity function, i.e.,
\begin{equation}
E\left(c_{A}\right)=\int_{A}\lambda\left(x\right)\mbox{d}x\label{eq:expectation-vs-intensity} \;,
\end{equation}
where the integral is computed over area $A$. A slightly different viewpoint is given by the \emph{density function,
}which is a normalised version of the intensity function, defined as
\begin{equation}
\bar{\lambda}\left(x\right)=\frac{\lambda\left(x\right)}{\int_{\Omega}\lambda\left(x'\right)\mbox{d}x'}\label{eq:density-function} \;,
\end{equation}
where the integral in the denominator is over the \emph{observation
window $\Omega$, }which in our case corresponds to the monitor (we
cannot observe points outside of the observation window). The density
function integrates to 1 over the observation window and represents
a probability density: If we now define a random variable $z_{A}$
that is equal to one, when a (small) area $A$ contains one point and 0 otherwise,
we obtain 
\begin{equation}
p(z_{A}=1)=\int_{A}\bar{\lambda}\left(x\right)\mbox{d}x=\bar{\lambda}\left(x_{A}\right)\mbox{d}A\label{eq:density-vs-prob} \;,
\end{equation}
where $x_{A}$ is the center of area $A$ and $\mbox{d}A$ its area%
\footnote{If $A$ is small then $\bar{\lambda}\left(x\right)$ will be approximately
constant over $A$, and the integral simplifies to $\bar{\lambda}(x_{A})$
times the volume%
}.

{\sl Second-order properties.} The first-order properties inform us about how many points can be
expected to find in an area, or, in the normalized version, whether
we can expect to find a point at all. Second-order properties tell
us about \emph{interaction between areas: }whether for example we
are more likely to find a point in area $A$ if there is a point in
area $B$.

In the case of the point process ({Fig.~\ref{fig:illustration-intensity}}),
the points are generated independently and do not interact in any
way, so that knowing the location of one point tells us nothing about
where the other ones will be. As shown in the manuscript, this is
not so with fixation locations, which tend to cluster at certain distances. 

The second-order statistics of a point process capture such trends,
and one way to describe the second-order statistics is to use the
pair correlation function. The pair correlation function is derived
from the \emph{pair density function} $\rho\left(x_{A},x_{B}\right)$,
which gives the probability of finding points at \emph{both }location
$x_{A}$ and location $x_{B}$. Let us consider two random variables
$z_{A}$ and $z_{B}$, which are equal to 1, if there are points in their
respective areas $A$ and $B$, and 0 otherwise (again we assume that
the areas are small). The probability that $z_{A}=1$ and that $z_{B}=1$
individually is given by the density function, Eq.~(\ref{eq:density-function}).
The probability that \emph{both} are equal to one is given by the pair density
function,
\begin{equation}
p(z_{A}=z_{B}=1)=\int_{A}\int_{B}\rho\left(x,x'\right)\mbox{d}x\mbox{d}x'=\rho\left(x_{A},x_{B}\right)\mbox{d}A\mbox{d}B\label{eq:pair-density-function} \;.
\end{equation}
The pair density function already answers our question of whether
observing a point in $A$ makes it more likely to see one in $B$,
and vice-versa. If points are completely independent, then the resulting pair density is given by
\begin{equation}
p(z_{A}=z_{B}=1)=\lb\left(x_{A}\right)\lb\left(x_{B}\right)\mbox{d}A\mbox{d}B\label{eq:pair-density-indep} \;.
\end{equation}

If the pair density function gives us a different result then an interaction is occuring. Therefore, if we take the ratio of the pair density, Eq.~(\ref{eq:pair-density-function}) to the product of the densities, we obtain a measurement of deviation from statistical independence, i.e.,
\begin{equation}
c\left(x,x'\right)=\frac{\rho\left(x,x'\right)}{\lb\left(x\right)\lb\left(x'\right)}\label{eq:dev-from-indep}
\end{equation}
The resulting object is, however, a complicated, four-dimensional (i.e., two dimensions for $x$ and two dimensions for $x'$) function and in practice it is preferable to use a summary measure, which is the
pair correlation function expressing how often pairs of points
are found at a distance of $\epsilon$ from each other. The pair correlation
function is explained informally in {Figure~\ref{fig:From-pdf-to-pcf}}. 

%-----------------
% Figure 9
%-----------------

More formally, the pair correlation function is just an average of
$c(x,x')$ for all pairs $x,x'$ that are separated by a distance
$r$, i.e.,
\begin{equation}
\rho\left(r\right)=\int_{x}\int_{x'\in\Omega|d(x,x')=r}c(x,x')\mbox{d}x\mbox{d}x'\label{eq:pair-correlation-function}
\end{equation}
In the above equation, the notation $x'\in\Omega|d(x,x')=r$ indicates that
we are integrating over the set of all points $x'$ that are on a
circle of radius $r$ around $x$ (but still in the observation window
$\Omega$). 

If we are to estimate $\rho\left(r\right)$ from data, we need an
estimate of the intensity function (as it appears as a correction
in Eq.~(\ref{eq:dev-from-indep})). In addition, since we have only
observed a discrete number of points, the estimated pair density function
can only be estimated by smoothing, which is why a kernel function
needs to be used. We refer readers to \cite{Illian2008} for details on pair correlation
functions. 

\end{appendix}

%-----------------------------------------------
\begin{figure}[p]
\begin{picture}(500,300)
%\put(0,380){\Large\bf Figures}
\put(-40,-10){\includegraphics[width=190mm]{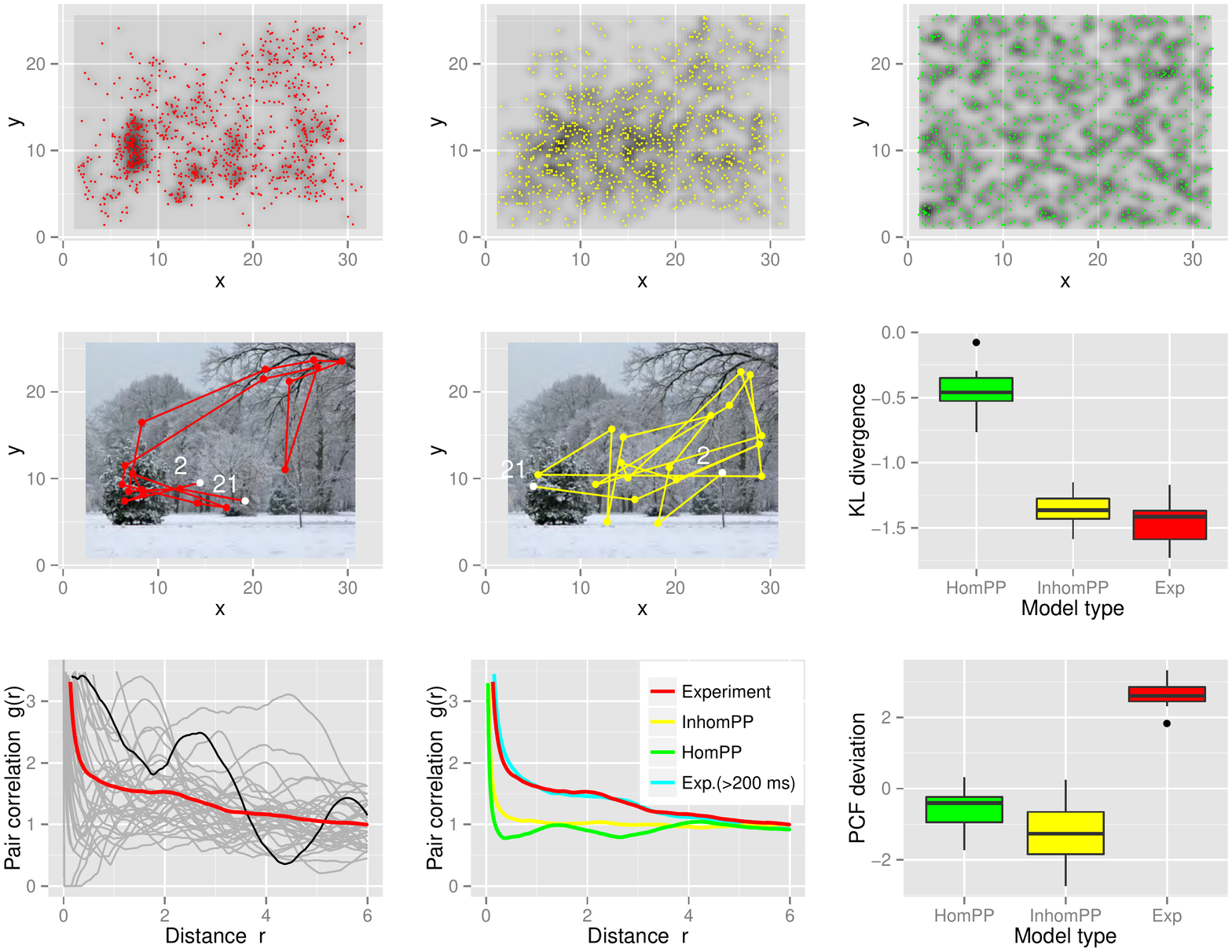}}
\put(2,348){\sf\LARGE a}
\put(155,348){\sf\LARGE b}
\put(310,348){\sf\LARGE c}
\put(2,226){\sf\LARGE d}
\put(155,226){\sf\LARGE e}
\put(310,226){\sf\LARGE f}
\put(2,109){\sf\LARGE g}
\put(155,109){\sf\LARGE h}
\put(310,109){\sf\LARGE i}
\end{picture}
\caption{\label{Fig_spatstat}
Analysis of pair correlation functions for experimental gaze sequences \re{(image set 1)} and for \re{computer-generated}surrogate data. {\bf (a)} Experimental data of gaze positions from human observers (red) and estimated intensity from kernel density estimate (gray levels) for image \#2. {\bf (b,  c)} Realizations of gaze positions generated by inhomogeneous and homogeneous point processes, resp. {\bf (d, e)} Typical single-trial fixation sequences from experiment (red) and inhomogeneous point process (yellow). {\bf (f)} Kullback-Leibler divergence (KLD) indicates that the inhomogeneous point process approximates the experimental 2D density of gaze positions. {\bf (g)} Pair correlation functions (PCFs) for experimental data (single trials: light gray; single trial from (d): black; averaged over trials: red). {\bf (h)} Mean PCFs for experimental data, inhomogeneous and homogeneous poisson process. {\bf (i)} The PCF deviation shows that the experimental data are spatially correlated, while the two surrogate datasets fail to reproduce this statistical pattern. }
\end{figure}%----------------------------------------------------

%-----------------------------------------------
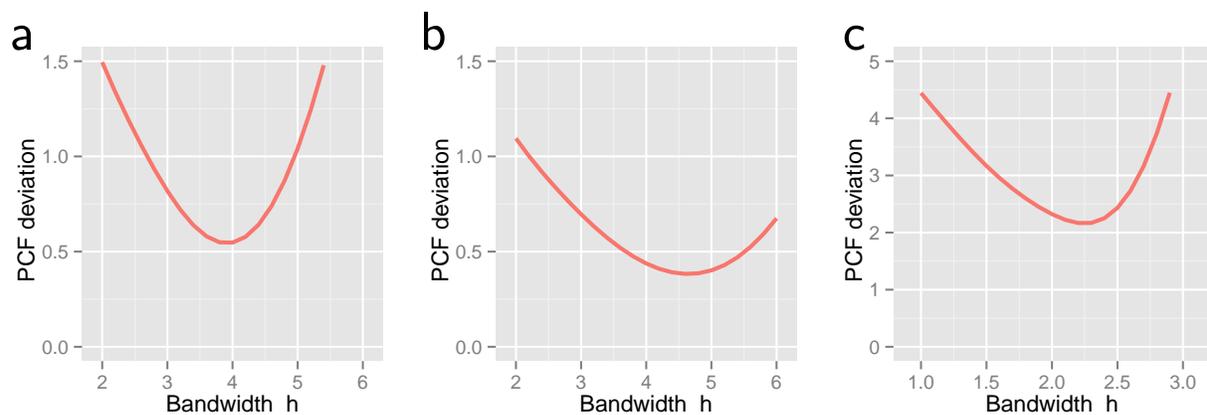
\begin{figure}[t]
\begin{picture}(500,150)
\put(-5,-5){\includegraphics[width=165mm]{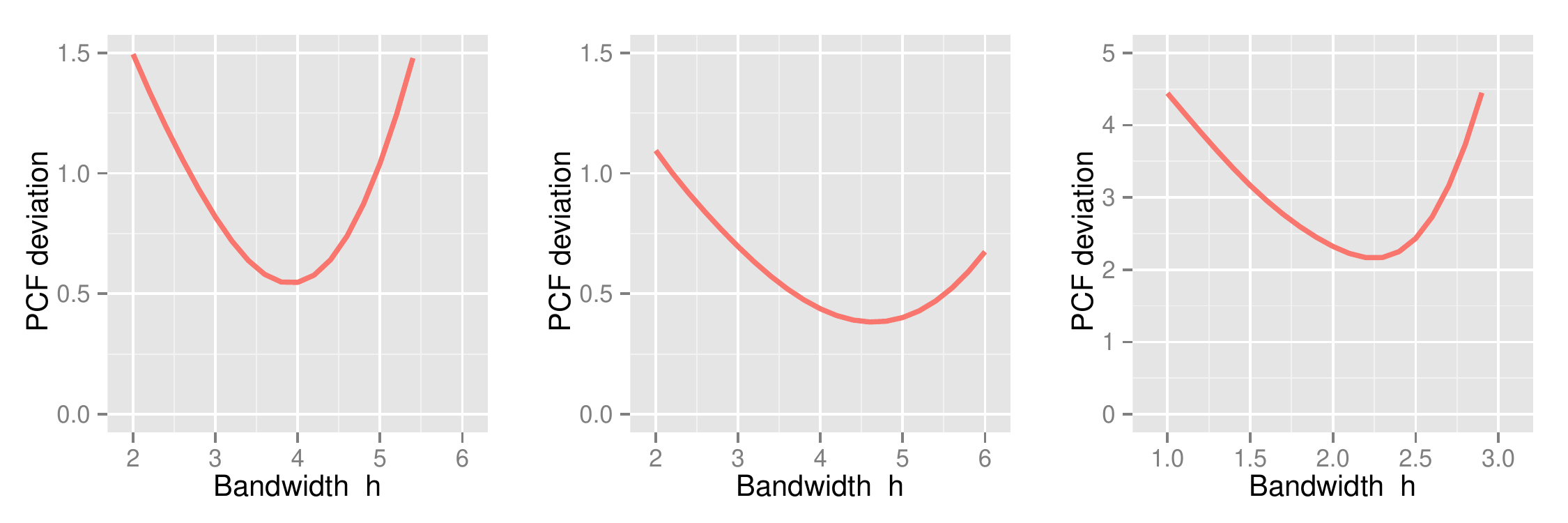}}
\put(0,140){\sf\LARGE a}
\put(155,140){\sf\LARGE b}
\put(315,140){\sf\LARGE c}
\end{picture}
\caption{\label{Fig_BW-PCF} Optimal bandwidth parameter for inhomogeneous pair correlation function (PCF) \re{for three different image sets}. For simulated data from the inhomogeneous point process, the PCF deviation $\Delta_g$, Eq.~(5), was computed as a function of the bandwidth $h$ \re{for the underlying kernel density estimate. The optimal bandwidth corresponds to the position of the minimum of $\Delta_g$. {\bf (a)} For image set 1, $\hat{h}_1=4.0^\circ$. {\bf (b)} For image set 2, $\hat{h}_2=4.6^\circ$.  {\bf (c)} For the Le Meur dataset, $\hat{h}_3=2.2^\circ$, due to a much smaller presentation display compared with our experiments.}}
\end{figure}%----------------------------------------------------

%-----------------------------------------------
\begin{figure}[t]
\begin{picture}(500,260)
\put(-40,-60){\includegraphics[width=193mm]{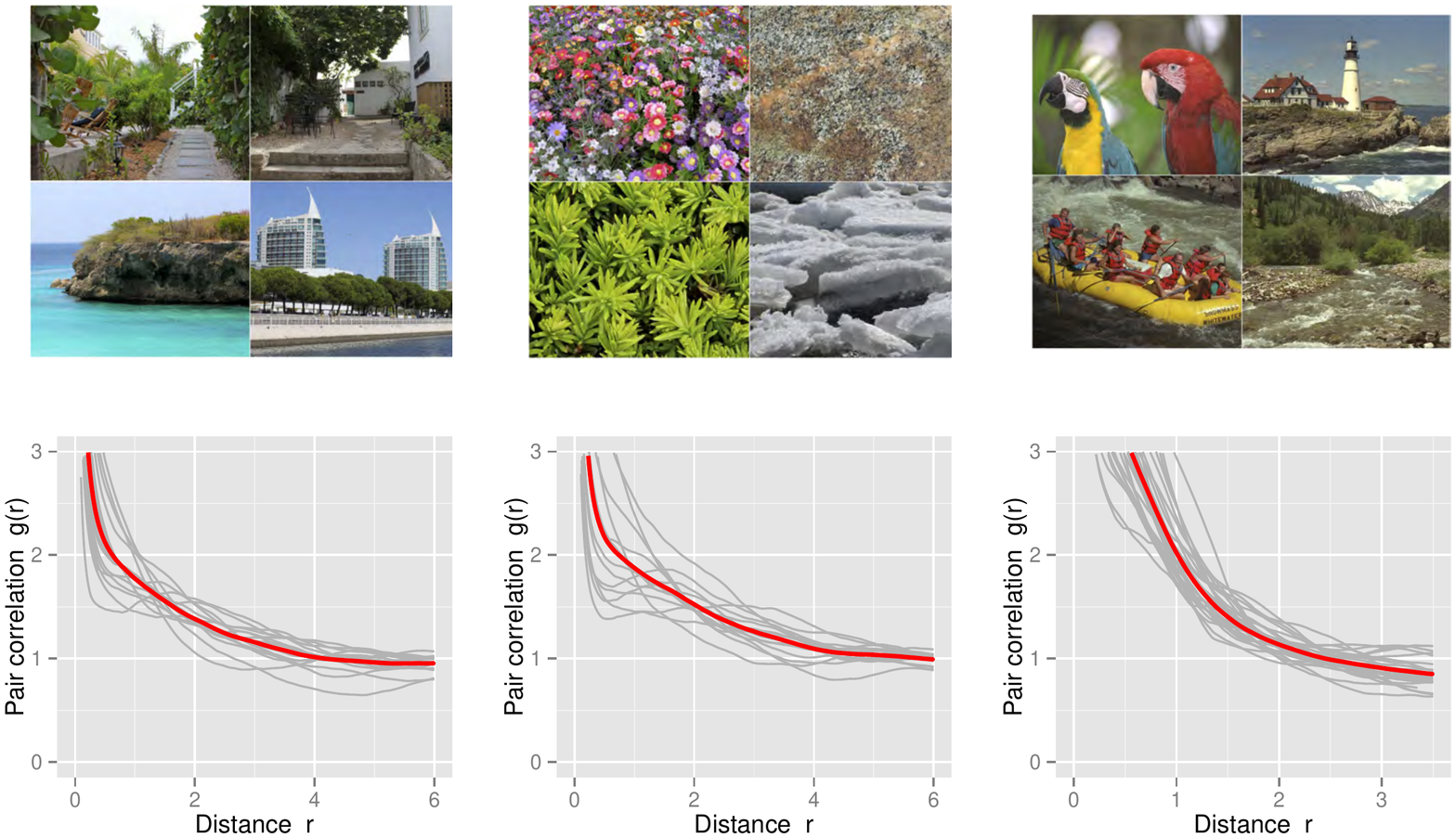}}
\put(0,255){\sf\LARGE a}
\put(155,255){\sf\LARGE b}
\put(315,255){\sf\LARGE c}
\put(0,120){\sf\LARGE d}
\put(155,120){\sf\LARGE e}
\put(315,120){\sf\LARGE f}
\end{picture}
\caption{\label{Fig_datasets}
\re{Comparison of the mean pair correlation functions for three different image sets: (a) Natural object-based scenes. (b) Abstract natural patterns. (c) Natural scenes from the Le Meur et al.~dataset. Mean pair correlation function are similar for object-based (d) and abstract (e) natural scenes. The presentation of images on a smaller display in the Le Meur dataset compared to our experiments (f) resulted in a smaller length scale of spatial clustering.}}
\end{figure}%----------------------------------------------------

%-----------------------------------------------
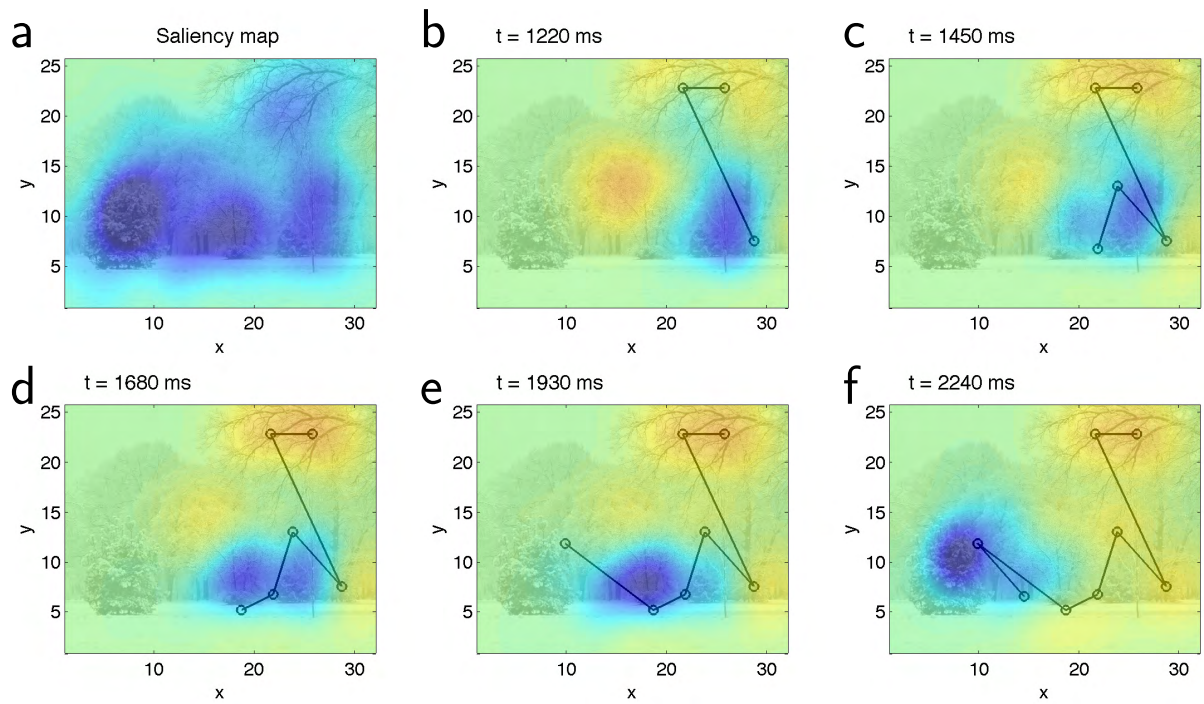
\begin{figure}[t]
\begin{picture}(500,250)
\put(-52,-15){\includegraphics[width=195mm]{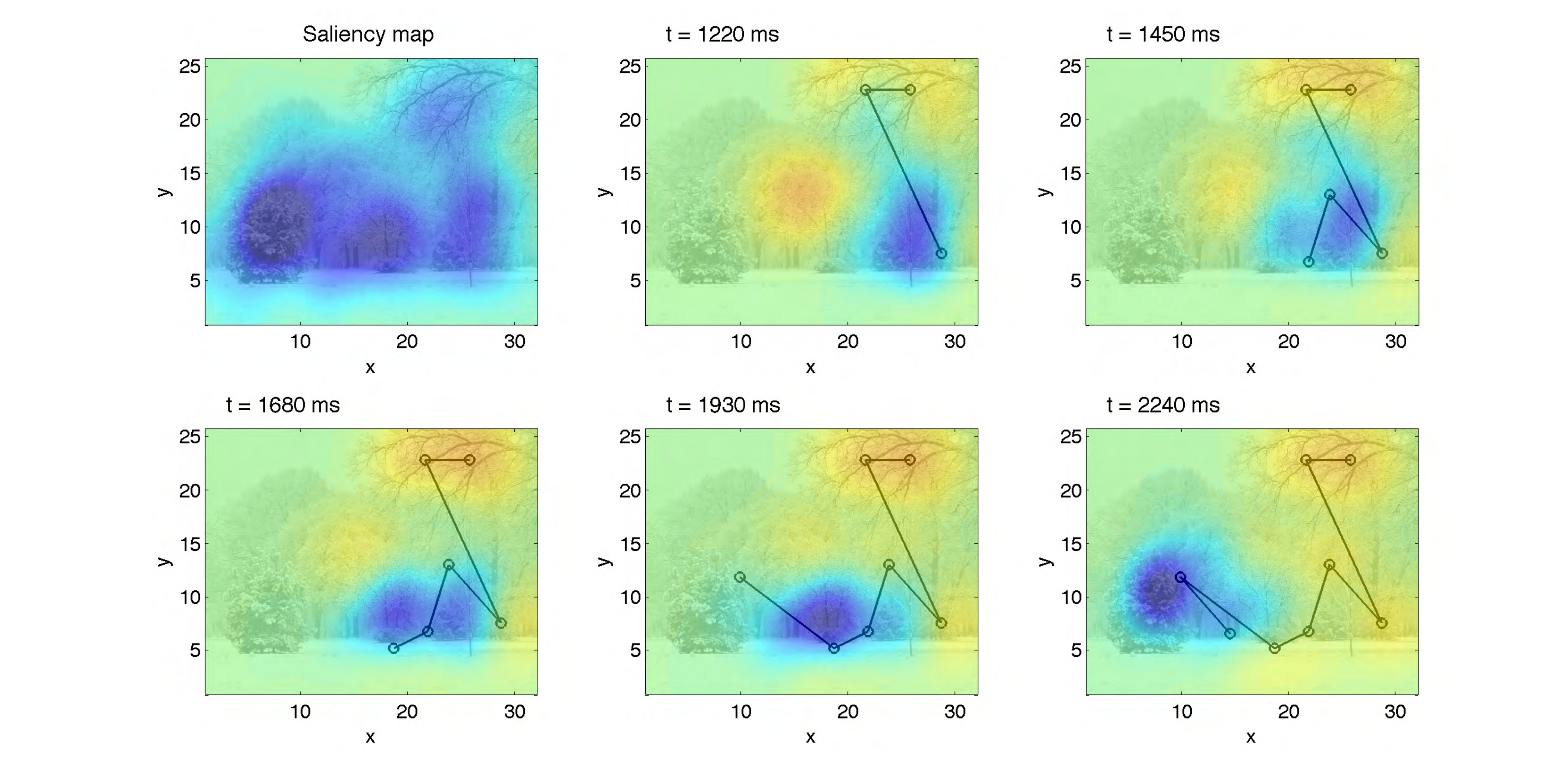}}
\put(0,245){\sf\LARGE a}
\put(155,245){\sf\LARGE b}
\put(315,245){\sf\LARGE c}
\put(0,110){\sf\LARGE d}
\put(155,110){\sf\LARGE e}
\put(315,110){\sf\LARGE f}
\end{picture}
\caption{\label{Fig_simexam}
Illustration of a simulated sequence gaze positions and the activation dynamics of the model. {\bf (a)} The density of gaze positions (empirical saliency) is used as a proxy for a computed saliency map that drives activation in the attention map. {\bf (b-f)} Sequence of snapshots of the potential (blue=low, yellow=high).}
\end{figure}%----------------------------------------------------

%-----------------------------------------------
\begin{figure}[p]
\begin{picture}(500,300)
\put(0,-10){\includegraphics[width=160mm]{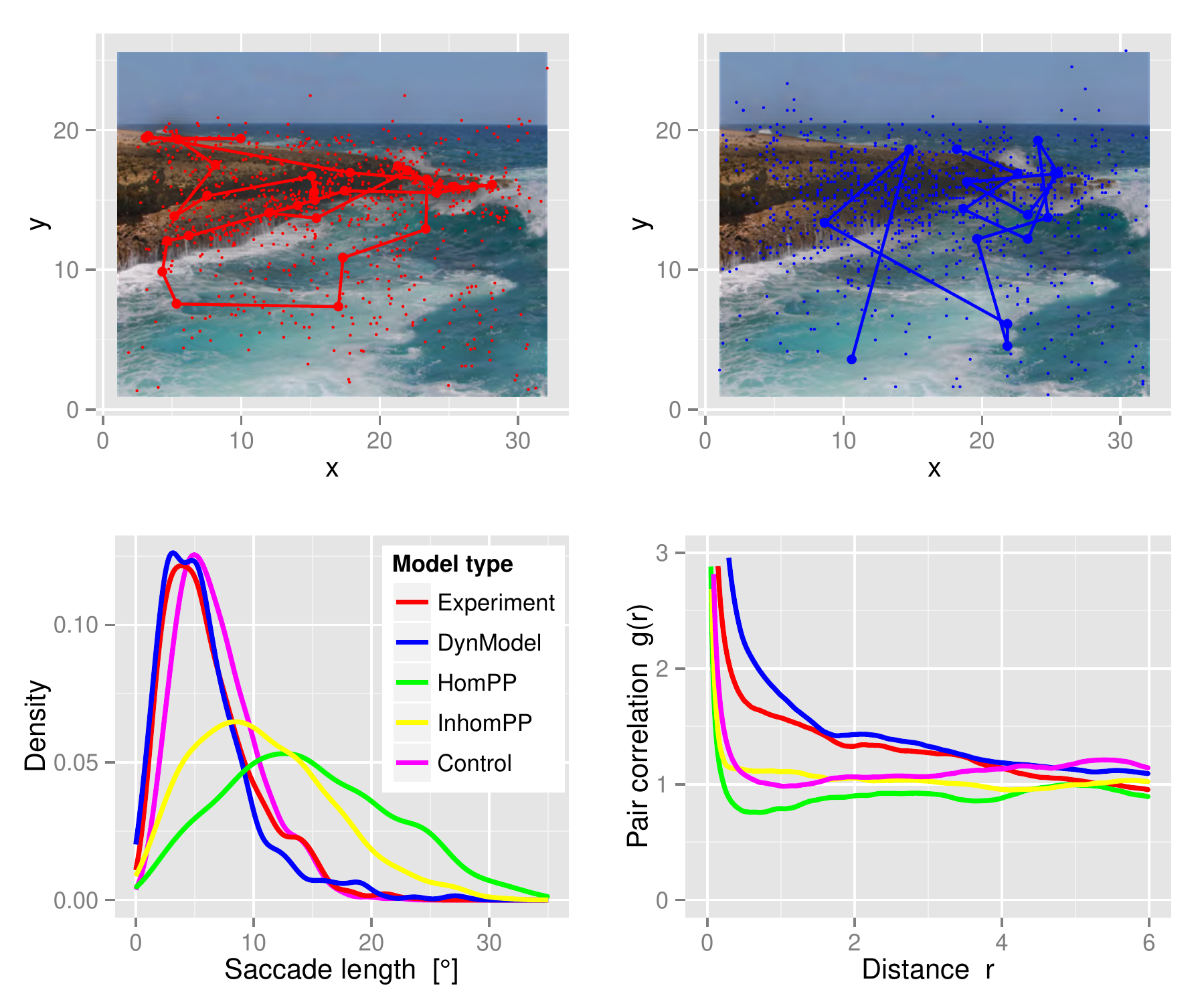}}
\put(5,355){\sf\LARGE a}
\put(233,355){\sf\LARGE b}
\put(5,165){\sf\LARGE c}
\put(233,165){\sf\LARGE d}
\end{picture}
\caption{\label{Fig_modelanalysis}
Distribution of saccade lengths and pair correlation functions from model simulations \re{(image \#6).} {\bf (a)} Experimental distribution of gaze positions (red dots) and a representative sample trial (red lines). {\bf (b)} Corresponding plot of simulated data obtained from our dynamical model (blue dots). A single-trial simulation is highlighted (blue line). {\bf (c)} Distributions of saccade lengths for experimental data (red), dynamical model (blue), homogeneous point process (green), inhomogeneous point process (yellow), and a statistical control model (magenta). {\bf (d)} Pair correlation functions for the different models. The dynamical model (blue line) produces spatial correlations similar to the experimental data (red line). }
\end{figure}
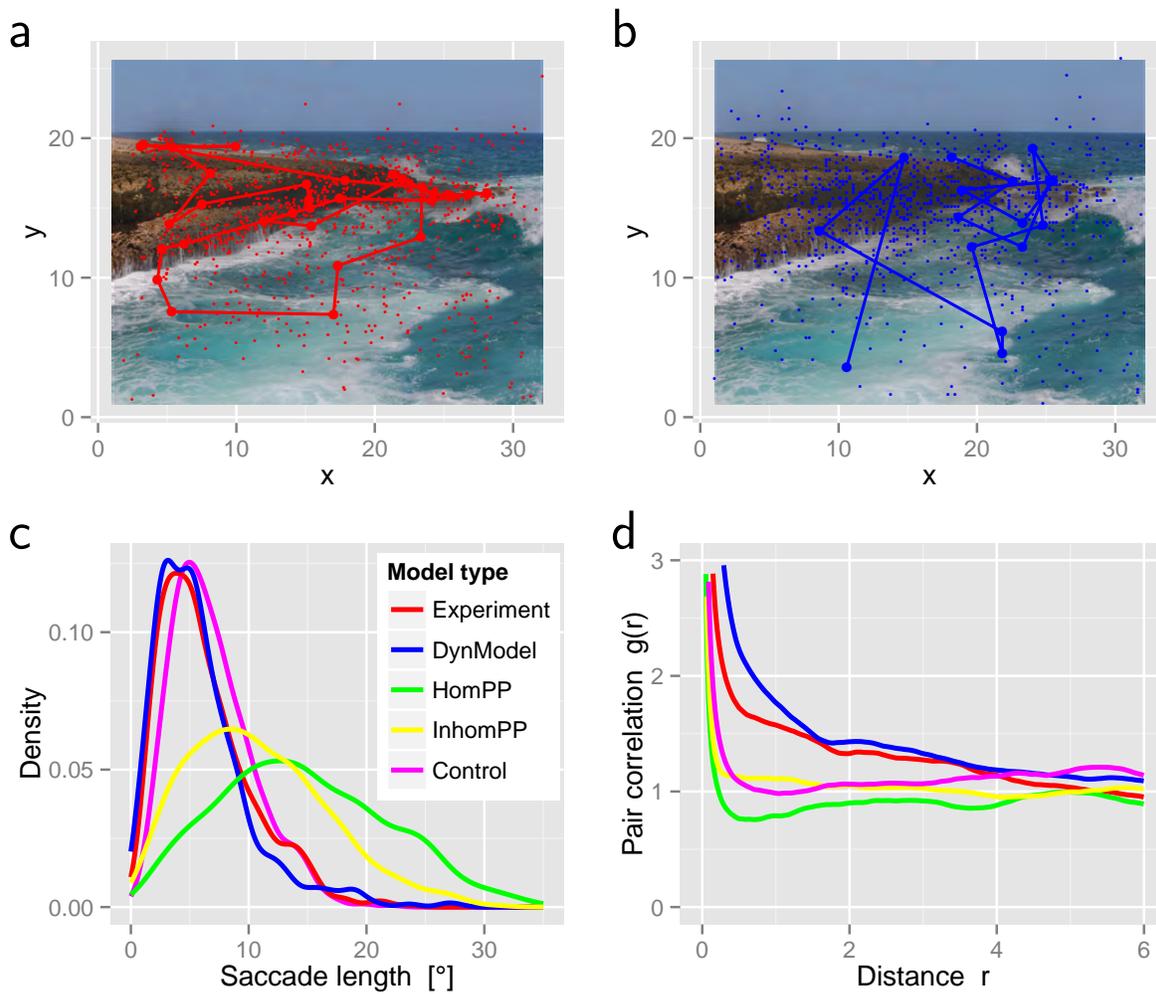%----------------------------------------------------

%-----------------------------------------------
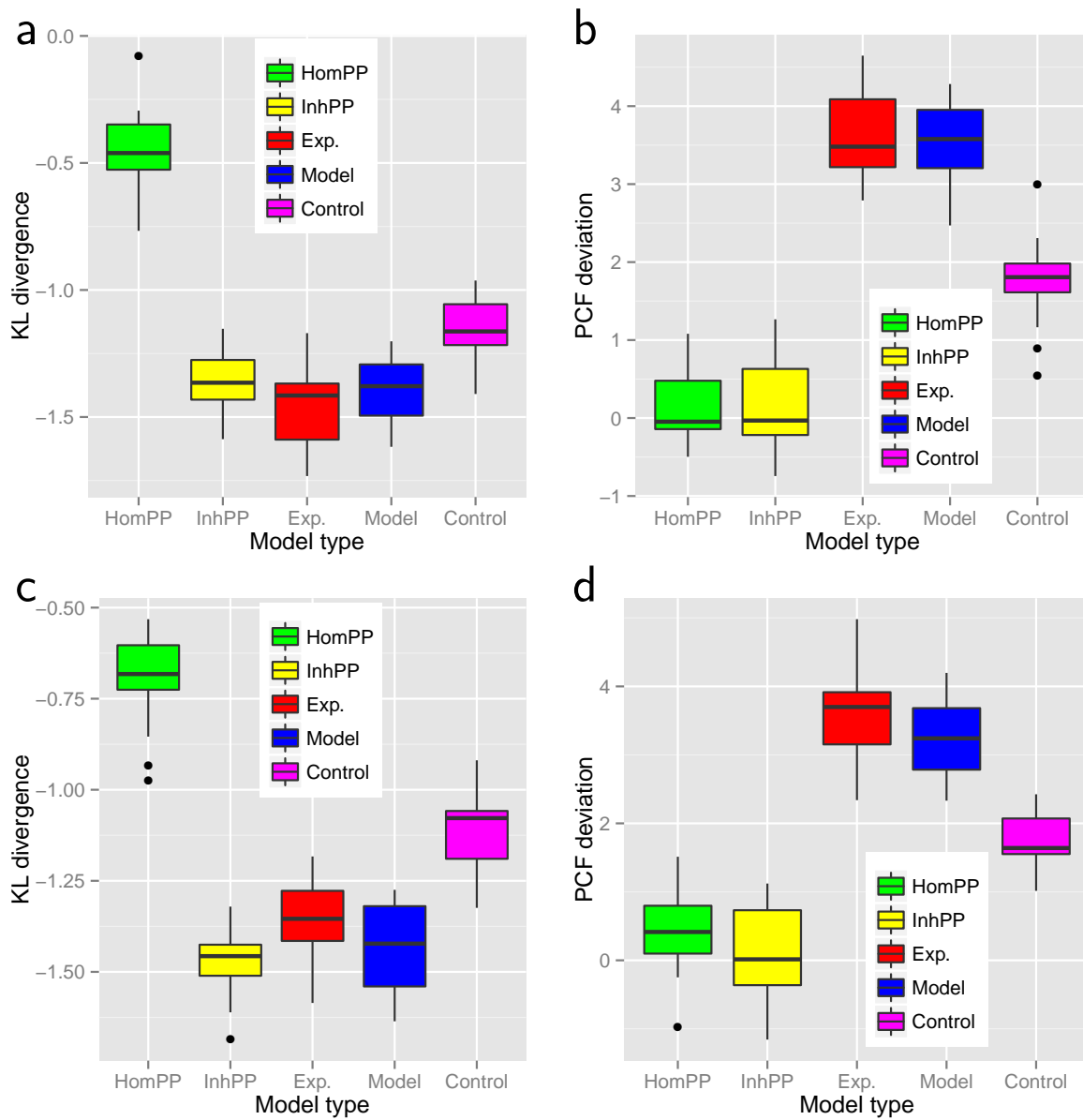
\begin{figure}[p]
\begin{picture}(500,405)
\put(-5,-10){\includegraphics[width=160mm]{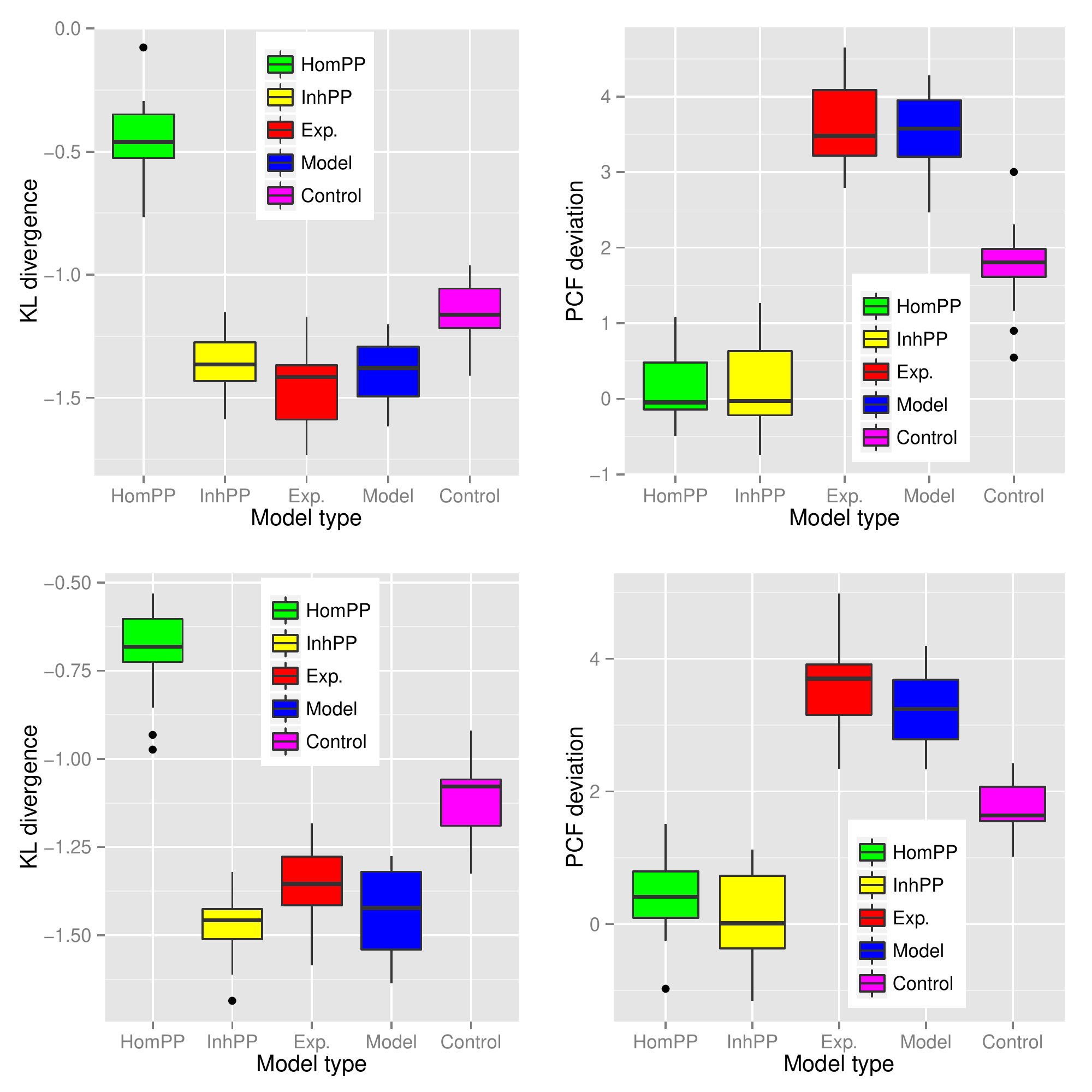}}
\put(5,430){\sf\LARGE a}
\put(230,430){\sf\LARGE b}
\put(5,205){\sf\LARGE c}
\put(230,205){\sf\LARGE d}
\end{picture}
\caption{\label{Fig_ModelResults}
Model predictions on images not used for parameter estimation. \re{Predicted data were generated from the dynamical model for the 10 images of image set 1 not used for parameter estimation (a,b) and for all 15 images from image set 2 (c,d).} (a) \re{Modell simulations of the KLD for experiments on image set 1 (10 images not used for model parameter estimation) and dynamical model and three different statistical models. For the experimental data, a split-half procedure was applied to compute KLD.} (b) Corresponding PCF deviations for the same model-generated and experimental data on image set 1. \re{(c) KLD measures for image set 2. (d) PCF deviation for image set 2.}}
\end{figure}
%----------------------------------------------------

%-----------------------------------------------
\begin{figure}[t]
\begin{picture}(500,300)
\put(20,-10){\includegraphics[width=140mm]{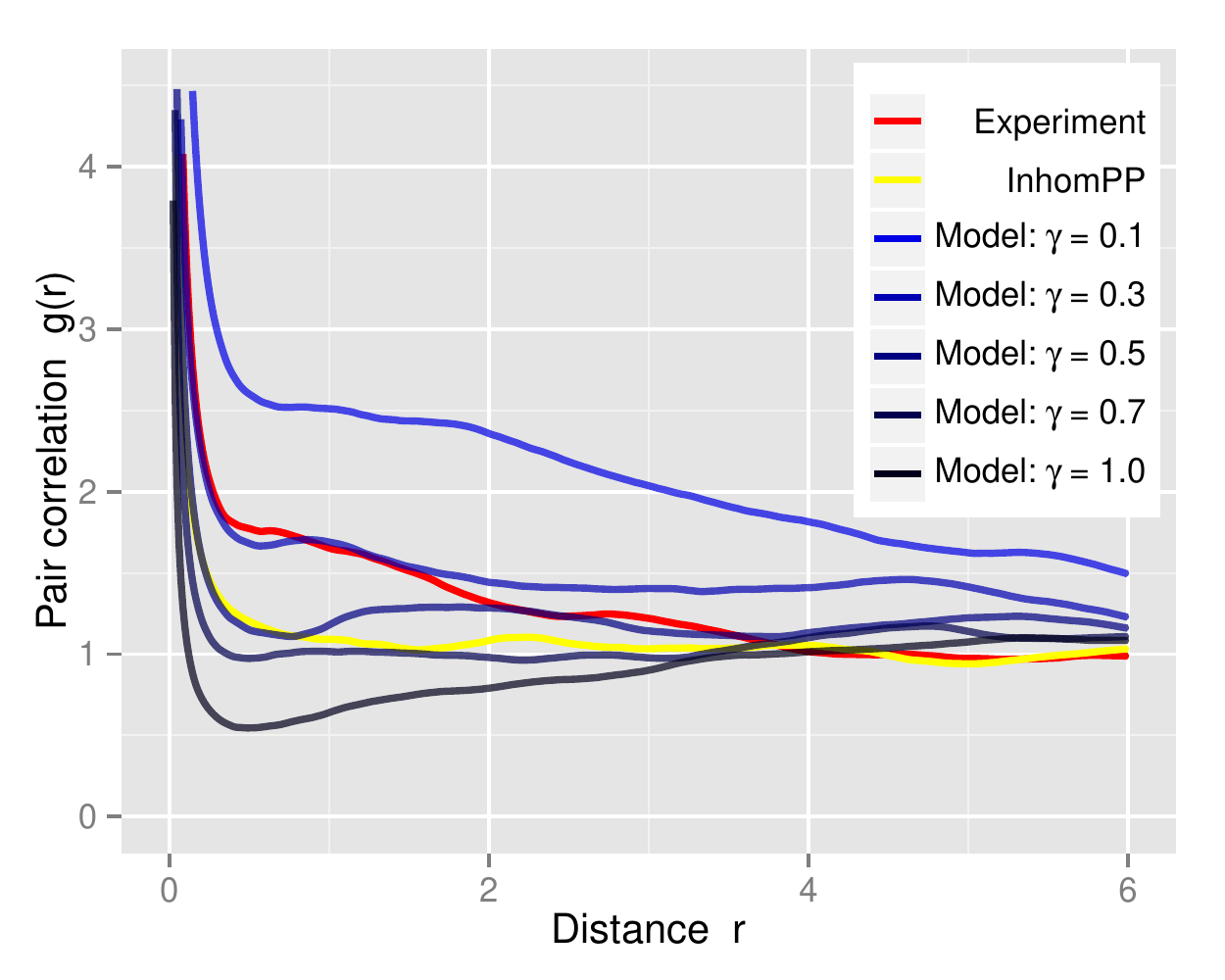}}
\end{picture}
\caption{\label{Fig_QA-Mod} Pair correlation function obtained from simulations for different values of $\gamma$ (blue lines) in comparison to the experimentally observed PCF (red line) and the result for the inhomogeneous point process (yellow line). All simulations were carried out for image \#1 of our data set.
}
\end{figure}%----------------------------------------------------

%-----------------------------------------------
\begin{figure}[h]
\begin{picture}(500,150)
\put(0,-10){\includegraphics[width=160mm]{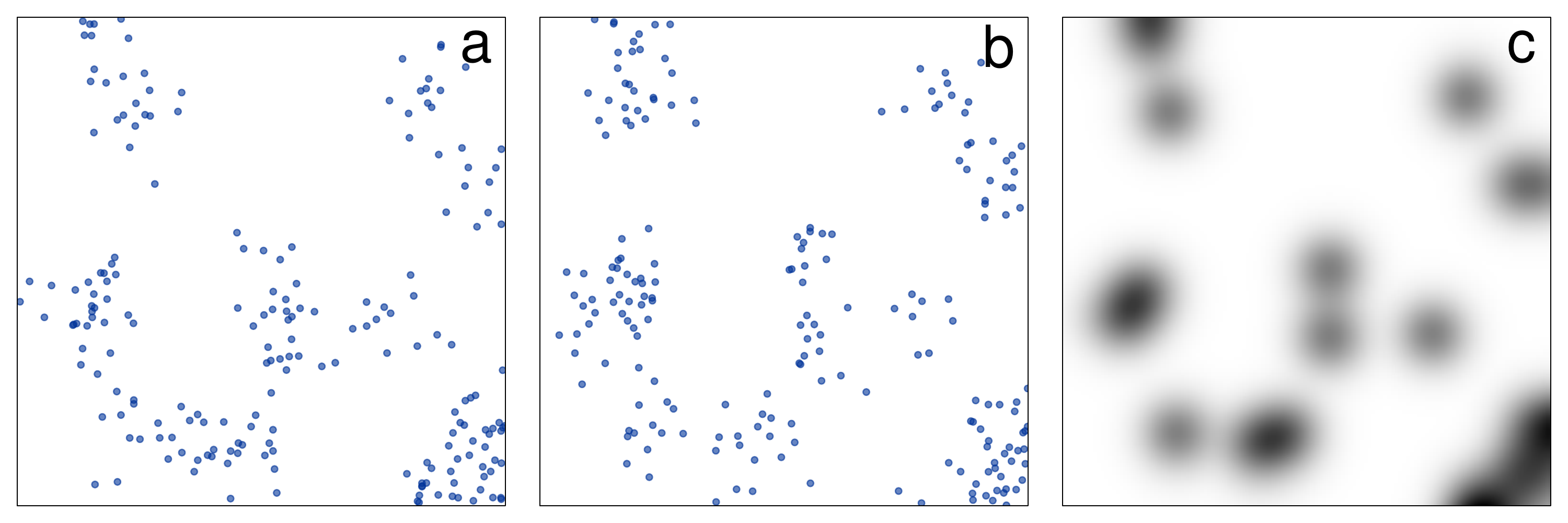}}
\end{picture}
\caption{\label{fig:illustration-intensity}
First-order properties of point processes. \textbf{(a}, \textbf{b)}
Two samples from a point process \textbf{(c)} The intensity of the
point process, $\lambda\left(x\right)$, which corresponds to the
expected number of points to be found in a small circle around location
$x$. Dark regions indicate high intensity (density).
}
\end{figure}%----------------------------------------------------

%-----------------------------------------------
\begin{figure}[t]
\begin{picture}(500,250)
\put(30,-50){\includegraphics[width=120mm]{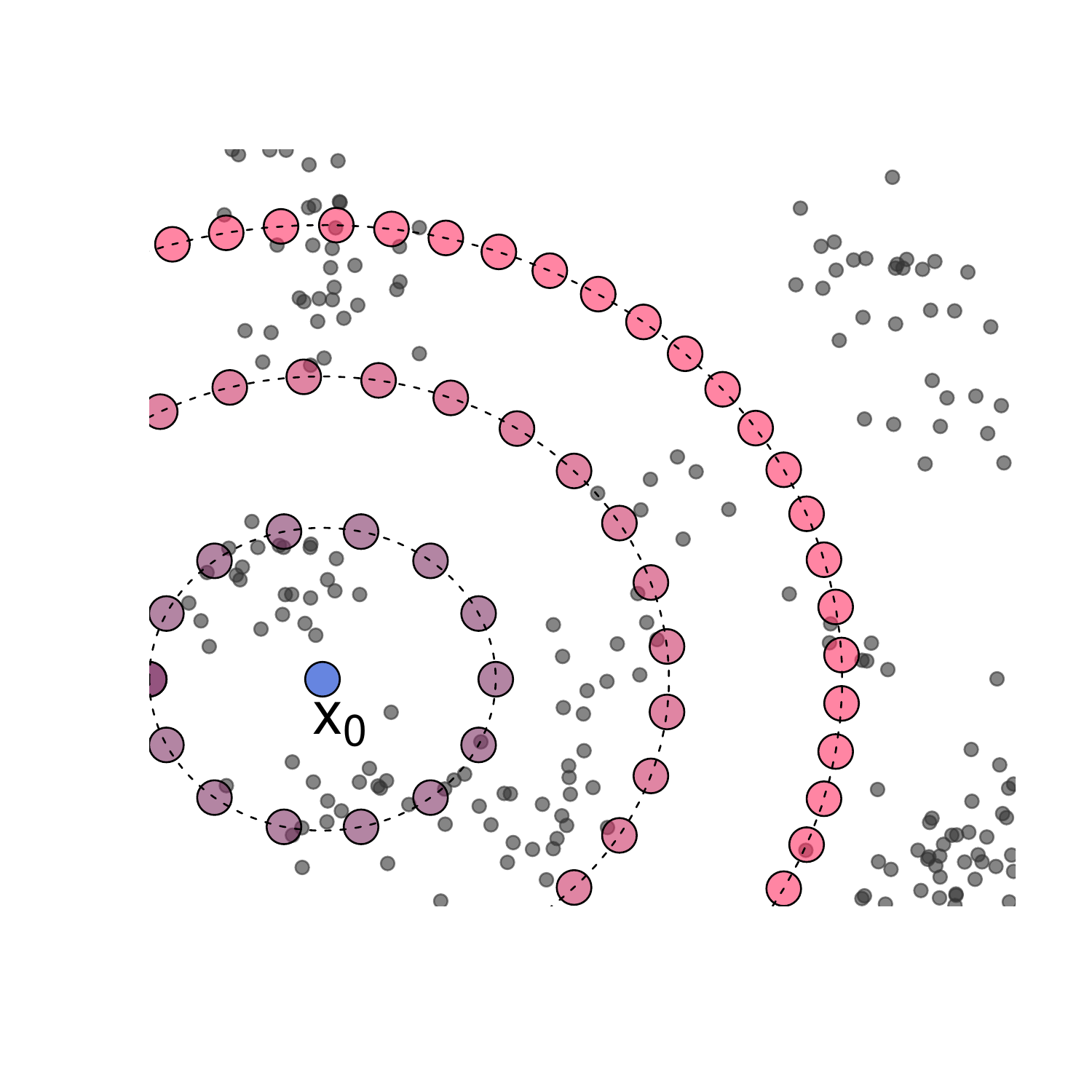}}
\end{picture}
\caption{\label{fig:From-pdf-to-pcf}
From the pair density function to the pair correlation function. The
pair density function $\rho\left(x_{A},x_{B}\right)$ describes the
probability of finding points at both $x_{A}$ and $x_{B}$ in a sample
from the point process. As such it is a four-dimensional function,
and hard to estimate and visualise. The pair correlation function
(PCF) is a useful summary. To compute the raw pcf, we pick an initial
location $x_{0}$ (circle) and look at the probability of finding
a point both at $x_{0}$ and in locations at a distance $\epsilon$
from $x_{0}$ (first array of circles around $x_{0}$). We do this
for various distances (other arrays of circles) to compute the probability
of finding pairs as a function of $\epsilon$. Finally we average
over all possible locations $x_{0}$, to obtain the pair correlation
function. The pair correlation function therefore expresses how likely
we are to find two points at a distance $\epsilon$ from each other.
}
\end{figure}%----------------------------------------------------

\end{document}